# Room temperature spin-layer locking of exciton-polariton nonlinearities

Jiaxin Zhao[1,†], Antonio Fieramosca[2,†,*], Kevin Dini[1,†], Qiuyu Shang[1], Ruiqi Bao[1], Yuan Luo[3], Kaijun Shen[4], Yang Zhao[4], Rui Su[1], Jesús Zúñiga Pérez[1], Weibo Gao[1], Vincenzo Ardizzone[2], Daniele Sanvitto[2], Qihua Xiong[3,5,6,7,*], Timothy C. H. Liew[1,*]

**Affiliations**
[1]Division of Physics and Applied Physics, School of Physical and Mathematical Sciences, Nanyang Technological University, Singapore 637371
[2]CNR NANOTEC Institute of Nanotechnology, via Monteroni, 73100 Lecce, Italy
[3]State Key Laboratory of Low-Dimensional Quantum Physics and Department of Physics, Tsinghua University, Beijing 100084, P.R. China
[4]School of Materials Science and Engineering, Nanyang Technological University, Singapore 639798, Singapore
[5]Frontier Science Center for Quantum Information, Beijing 100084, P.R. China
[6]Beijing Academy of Quantum Information Sciences, Beijing 100193, P.R. China
[7]Beijing Innovation Center for Future Chips, Tsinghua University, Beijing 100084, P.R. China.

†These authors contributed equally to this work

*To whom correspondence should be addressed. Emails: tchliew@gmail.com, Qihua_xiong@tsinghua.edu.cn and antonio.fieramosca@gmail.com

**Abstract**

Spintronics is pivotal in the infrastructure of modern information technology. Despite intensive research efforts dedicated to the exploration of innovative device architectures, establishing a materials platform conducive to spin optronics, particularly one that operates effectively at ambient temperatures, continues to represent a significant challenge. Recent advancements in transition metal dichalcogenides (TMDs) have unveiled exceptional optical and electronic characteristics, opened up new opportunities, and provided a unique platform for exploring light-matter interactions under the strong coupling regime. The exploitation of exciton-polaritons, with their peculiar hybrid light-matter properties, for the development of spintronic customizable devices that enhance both the information capacity and functionality at ambient temperatures is often suggested as a promising route. However, although TMD polaritons have shown promising potential, the microscopic mechanisms leading to nonlinearities in TMD polaritons are complex and their spin-anisotropy, a crucial requirement for many proposed polaritonic devices, has been missing. Here, we demonstrate the absence of spin-anisotropic interaction in a monolayer $WS_2$ microcavity (at room temperature) and show how spin-dependent interactions can be controlled and spin anisotropy recovered by engineering double $WS_2$ layer structures with varied interlayer spacing. We attribute this phenomenon to a distinctive feature in exciton-polariton physics: layer-dependent polariton-phonon coupling. We use theoretical calculations of the phonon electrostatic potentials finding a drastically different coupling strength for single and double monolayer samples and discuss qualitatively how this explains the observed spin-anisotropic response. This is further consistent with experiments on multi $WS_2$ layer samples and the identification of a critical separation distance, above which an effective single monolayer spin-anisotropic response is recovered, both in experiment and theory. Our work lays the groundwork for the development of spin-optronic polaritonic devices at room temperature.

**MAIN TEXT**

Spintronics has demonstrated significant potential in advancing next-generation memory and computing technologies, with spin-based architectures forming a cornerstone[1]. Excitons in semiconducting materials have been identified as a promising candidate for future spintronic devices in a recent roadmap due to their ability to be encoded or read both optically and electrically[2]. However, some of the most advanced proof-of-concept demonstrations of exciton spin-based logic, such as those based on exciton-polaritons[3], have seen limited development given their cryogenic cooling requirements. Indeed, exciton-polaritons have been traditionally studied in III-V materials, where the small exciton binding energy places a need for temperatures around a few Kelvin for their existence. The development of wider bandgap materials has enabled exciton-polaritons to reach room temperature, with high-speed (picosecond or lower) switching devices being implemented in perovskite[4] and organic microcavities[5]. However, these devices have not explicitly exploited the spin degree of freedom. A good spintronic platform should also exhibit good spin coherence, which is somewhat limited in many materials where large anisotropy (especially in perovskite) would cause rapid spin relaxation[6].

The demonstrations of tightly bound excitons and robust oscillator strengths in TMD monolayers enables the persistence of TMD polaritons even at room temperatures[7], paving the way for practical optical and polaritonic device applications. Monolayer TMDs, characterized by time-reversal symmetry, strong spin-orbit coupling, and broken crystal inversion symmetry, exhibit pronounced spin-valley coupling. The opposite spins (+1 or -1) are unambiguously associated with the different valleys in the momentum space ($K$ or $K'$), allowing for the direct optical initialization of exciton spin in these non-equivalent valleys with opposite circular polarizations ($\sigma+$ or $\sigma-$). The manipulation of valley degrees of freedom in TMD monolayers via helicity-controlled excitation of light has reinvigorated interest in valleytronics, suggesting the use of monolayers for encoding information in electronic valley degrees of freedom. When strongly coupled with microcavities, monolayer TMD polaritons preserve the valley pseudospin and inherit the properties from the photonic component, facilitating long-distance information transfer and modified relaxation dynamics with enhanced valley coherence[8-10]. Recent demonstrations, such as manipulation of valley coherence[11], optical valley Hall effects[12], and valley-selective optical Stark effect[13], further underscore the potential of monolayer TMD polaritons. Furthermore, the exploration and manipulation of strong nonlinearities inherited from the exciton component lie at the core of groundbreaking polaritons physics and extensive research has uncovered mechanisms of enhancing effective polariton-polariton interactions, by using for instance Fermi polarons[14], trion polaritons, excited-state Rydberg polaritons[15], and Moiré potentials[16] from twisted heterostructures. Therefore, TMDs turned out to be an excellent platform for studying and implementing polariton physics.

The use of exciton-polaritons in spintronic device proposals has been very highly dependent on the spin anisotropy of interactions. In cryogenically cooled III-V materials, polaritons with parallel spins can repel at least an order of magnitude stronger than polaritons with antiparallel spins. Turing complete computational architectures proposed with exciton-polaritons rely on this effect[17,18], together with a range of relevant fundamental effects, including: the spin Meissner effect[19], self-induced Larmor precession for magnetometry[20], and polarization multistability for spin memories[21]. In the room temperature polariton framework, the same spin anisotropy of interactions has been evidenced in bulk perovskites[22-24] and, only reported in TMD-based superlattice microcavities very recently[25]. At the state of the art, the complex interplay between different phenomena occurring in single or multiple monolayers working under strong coupling regime has not been thoroughly investigated, thus limiting the development of this promising platform for polariton spintronics.

In the present work, we propose and experimentally demonstrate, using polarization resolved nonlinear spectroscopy, that the physics of structures with multiple TMD layers is fundamentally different than that of single TMD layers for exciton-polaritons in microcavities. In particular, the strong light-matter interaction that forms exciton-polaritons enforces that the lowest energy (lower branch) polaritons must have the same phase in the different TMD layers and we propose that this reduces their coupling to optical phonons. Indeed, while optical phonons coupling to a single layer are allowed to have a wide range of momenta in the growth direction, coupling to multiple layers requires states with specific momentum. This effect, which is largely independent of TMD layer separation (at least before a critical value), is unexpected for electrons in multiple quantum wells or excitons in multiple TMD layers, and unique to exciton-polaritons. The result is that while the polaron interaction, which is independent of spin, may be dominant in single TMD layer microcavities, it can be suppressed in multi-TMD layered cavities allowing recovery of spin-dependent interactions attributed to phase space-filling or the exchange interaction.

**Results**

In our experiments, we used two different kinds of structures: one containing a single $WS_2$ monolayer (ML) and one containing two $WS_2$ layers (N2) with a $SiO_2$ layer in between serving as a spacer. As depicted in Fig. 1a, the active structures, whether monolayers or N2, are integrated into planar microcavities formed by two distributed Bragg reflectors (DBRs). For the single layer samples, the monolayer is transferred onto a single DBR ending with a $\lambda_{Exc}/4$ $SiO_2$ layer. A polymethyl methacrylate (PMMA) layer is then spin-coated to tune the energy of the cavity mode, followed by the transfer of the top DBR, comprising of 6.5 pairs, which completes the fabrication of the microcavity. In the case of the N2 structure, the samples are formed by a stack of two $WS_2$ layers, each separated by an $SiO_2$ layer varying in thickness from 4 nm to 50 nm, as schematically shown in Fig. 1 b and c. Absorption and photoluminescence (PL) emission spectra are reported in Fig. 1 d. Detailed information about the sample's fabrication is provided in the Methods section. All the N2 samples show an increase in the absorption and PL emission and good spatial uniformity as reported in Section 1 of Supplementary Materials (SM) (Fig. S1).

The strong coupling between excitons and cavity photons has been effectively demonstrated by performing angle-resolved reflectivity measurements, as shown in the right part of Figure 1e, 1f, and 1g for the monolayer sample, and N2 samples with 6 nm and 30 nm spacers, respectively. Two distinct modes are observed: the higher-energy mode exhibits a flattened dispersion at small incidence angles, while the lower-energy mode's dispersion flattens at larger angles. This characteristic is indicative of a two-branch polariton dispersion, *i.e.*, the upper and lower polariton branch (UPB and LPB), confirming that all the samples are within the strong coupling regime. The left part of the figures shows simulation results, obtained by using a Transfer Matrix Method, which are in good agreement with the experimental data. Moreover, by fitting with a two coupled oscillator model, the Rabi splitting energy is found to be approximately 34 meV for monolayer and 52 meV for N2 samples. These values are in good agreement with previous reports[26,27], where the same TMD is employed in similar optical structures. All experiments were conducted at room temperature.

**Nonlinearities in the monolayer polaritons**

In the context of TMD-polaritons, numerous studies have reported the persistence of spin-valley locking, where the emission from valley polaritons can be coherently excited through off-resonance or near-resonance excitation[8-10]. Although these reports show the potential of valley-polaritons, they have not exploited the nonlinear regime. A strong polarization response of the PL emission is not sufficient for the realization of many proposed polariton-based devices, which instead require the exploitation of the nonlinear regime. Ideally, one hopes that polariton-polariton interactions are spin-anisotropic, where polaritons with the same spin strongly interact and particles with opposite spins exhibit weaker interactions. In cryogenically cooled samples, this is often a direct consequence of spin-selection rules in the Coulomb exchange interaction and phase space filling (PSF)[22,28-30]. It is worth noting that a deviation from this typical picture can occur if the nonlinear response of polaritons is driven by different microscopic mechanisms, such as dipolar interactions, screening of the exciton binding energy, or the presence of indirect exchange interactions mediated by biexciton states[31-

[34]. This is particularly relevant for TMD polaritons, where previous studies have suggested that interaction mechanisms involving trions or polarons may be dominant in TMD-based polariton systems[35,36].

Further, for a cryogenically cooled microcavity with a MoSe$_2$ monolayer a stronger interaction was visible for polaritons injected with opposite spins and a weaker interaction for polaritons with the same spin[34,35]. This is understood as a Feshbach resonance effect, where the formation of a biexciton polariton (composed of opposite spins) is dominant[36]. Therefore, we first investigated the polarization-dependent nonlinear response of our WS$_2$-monolayer microcavity samples by varying the exciton/photon contents of the polariton branches. This allows us to check the importance of the biexciton resonance on the spin-dependent character of polaritons. This variation in detuning was achieved by slightly adjusting the thickness of the PMMA layer, leading to a detuning variation from -5 meV down to -60 meV, as detailed in the Methods section. The angle-resolved reflectivity maps obtained from microcavities with the most positive (~-5 meV) and most negative (~-60 meV) detuning are presented in Figure 2a and b, respectively. The dispersions of all monolayer microcavities and the corresponding exciton/photon fractions of the LPB are included in the Section 1 of SM (Fig. S2). All samples are in the strong coupling regime, exhibiting a Rabi splitting of around 34 meV, in good agreement with previous works[37,38].

To investigate the polarization-dependent polariton-polariton interaction in monolayer, we conducted resonant nonlinear transmission measurements using a pulsed excitation (1kHz, 100 fs). A scheme of the employed setup is reported in Section 1 of SM (Fig. S3). The laser is directed at normal incidence with a small beam waist to avoid any potential influence from the transverse electric-transverse magnetic (TE-TM) splitting (Section 1 of SM, Fig. S4). Consistent with behaviors expected in a system of interacting particles, we observe a shift in the transmission peak toward higher energies with increasing density. We extract the blueshift of the ground state, quantifying the magnitude of the energy shifts relative to its value at the lowest density. The energy shifts as a function of the polariton density for samples with different detuning of approximately -5 meV, -15 meV, -35 meV and -60 meV are reported in Figure 2c, e, g and i, respectively. Orange dots represent experimental data using a linearly polarized excitation, while dark purple and dark blue dots indicate $\sigma+$ and $\sigma-$ polarization, respectively. Corresponding normalized transmission spectra at both low and high excitation densities for all the samples with different detuning are displayed in Figure 2d. f. h and j. In these figures, dots represent experimental data points, while lines indicate the results of Lorentzian fits.

As it is clear from the experimental data, the energy blueshift induced by the circularly polarized laser excitation (*i.e.*, when all polaritons have the same spin) is essentially the same as that measured with linearly polarized laser excitation (*i.e.*, when half of the excited polaritons have the same spin). The results indicate that there is a limited modulation in response to changes in the pump polarization, and no discernible indication of polarization-dependent interactions within ML polaritons, regardless of the detuning of the sample. These findings are in contrast with those obtained in GaAs and perovskites[22,28], indicating a strong deviation from traditional expectations[35]. Moreover, compared to the results obtained in $MoSe_2$ monolayer polaritons at cryogenic temperatures[35], where the presence of an intermediate biexciton state could justify the change in the sign of the interaction, our results do not indicate such behavior. Indeed, the biexciton states are relatively weak at room temperature and we conclude that they are ineffective at generating spin anisotropy in our case.

**Spin dependent interactions of polaritons based on N2**

The N2 samples consist of an arbitrarily twisted stack of two $WS_2$ monolayers, with a $SiO_2$ spacer layer in between. The thickness of the spacer is varied from ~4 nm up to 50 nm. This configuration ensures that interlayer hopping of electrons/holes at $\pm K$ valleys is minimal, owing to band misalignment in momentum space and the presence of spacer as hopping barriers[39]. Therefore, in all the N2 samples electronic coupling between the two layers is absent. It is important to note that for every SiO$_2$ thickness. the PMMA layer employed to tune the energy of the cavity mode is decreased in order to guarantee a comparable detuning for all the samples. The experimental dispersions and corresponding exciton/photon coefficients for all N2 samples are presented in Section 2 of SM (Fig. S5), while the TMM simulations of the dispersions for these samples are shown in Fig. S6. Table 1 reports the PMMA thickness employed in the simulation in order to match the energy of the polariton branches found experimentally. Moreover, the spatial distribution of the electromagnetic field of the LPB (normal incidence) across the microcavity structure is shown in Figs. S7 and S8. These simulations demonstrate that even at a large separation, both monolayers experience a significant electromagnetic field and that both TMD layers are involved in the exciton-polariton formation.

The transmission spectra under varying excitation densities for the N2 samples with different spacer thickness are reported in Figure 3a (6nm spacer and corresponding dispersion shown in Fig. 1e), Section 3 of SM Fig. S9 and S10 (all the other spacers). As illustrated in the bottom panel of Figure 3a, at low excitation densities, the transmission signals are indistinguishable between circular (blue curves) and linear (red curves) polarizations. However, as the excitation density increases, a more pronounced blueshift is observed in the circularly polarized excitation compared to that in the linearly polarized excitation. This different behavior underscores the effective spin-dependent nature of the interactions within the double TMD layer polaritons and it is remarkable given that no such behaviour is observed in the monolayer samples. Even in this case, we extract the blueshift of the ground state and plot it as a function of the polariton density. The power dependence for microcavities with spacer layers of approximately 6 nm, 20nm, 30 nm, and 40 nm is illustrated in Figures 3b, c, d and e, respectively. Additional N2 samples with a different spacer thickness are shown in Section 3 of SM (Fig. S9 and S10). In all samples, with the same excitation density, the energy blueshift observed under circularly polarized excitation (blue dots) consistently exceeds that under linearly polarized laser excitation (red dots). We used a linear fit of the power dependence data to estimate the slope in both the linear (L) and circular cases (C), as indicated by the red and dashed lines, respectively. The ratio between the two slopes, C/L, a measure of spin-anisotropy of interactions, as a function of the spacer thickness is shown in Fig. 4a. Specifically, we plot here the ratio obtained by using either a clockwise (blue points) or counterclockwise circular polarization (red points). The value obtained for the monolayer sample (C/L close to 1±0.15) is used as a reference. As can be seen, the C/L ratio is larger than one for the N2 samples. Although the exact ratio depends on the number of points taken in the fit for the blueshift data (Fig. 3), it fluctuates around an almost constant value for spacer thickness up to 30nm. At a spacer distance of 40nm, the effective spin-dependent interactions become weaker, behaving effectively the same as in the monolayer.

In order to explain the aforementioned results, it is important to first rule out some possibilities. First, a possible consideration could be differences in doping levels between the monolayer and N2 samples, which would change trion concentrations. To further investigate, we conducted PL mapping of the N2 sample without the top DBR, utilizing different bandpass filters to selectively capture emission from the exciton and trion complexes. As shown in Section 4 of SM (Fig. S11), the trion emission from both N2 and monolayer regions exhibits a comparable ratio, suggesting that the doping levels are comparable for all the fabricated samples. Second, all known mechanisms of interactions between layers (including the formation of interlayer excitons, interlayer trion complexes, biexciton coupling, electron/hole tunneling between layers, Förster energy coupling) are not expected for the large spacer distances. Additionally, had they been present, we should have seen a smooth monotonic reduction of their effect with spacer distance. In contrast, as shown in Fig. 4a, there is no smooth reduction in the C/L ratio as the spacer thickness increases, but rather a roughly constant value before a critical spacer thickness. Third, given that there is no interlayer exciton interaction, the specific mechanism of interaction in monolayers is less relevant (the exchange interaction and phase-space filling are discussed in the literature) for the present problem as they should all be as active in multilayered samples as monolayers. Fourth, we do not expect other polarization sensitive effects such as TE-TM splitting (negligible at normal incidence) and spin-layer valley locking to be playing a role and, again, if they did, they should be equally active in multilayered and monolayer samples. The present observations rather point to an unexpected fundamental difference between N2 and monolayer samples.

**Model of spin-dependent blueshifts in mono and N2 samples**

In order to model the experimental results reported above, we first considered how bosonic counting statistics might influence the spin dependence of polariton-polariton interactions, as discussed in Section 5 of SM. From this analysis, we found that the blueshift for circularly polarized excitation is larger than that for linearly polarized excitation (Fig. S12), even though we assumed equal spinor interaction constants for a single monolayer ($\alpha_2 = \alpha_1$, *i.e.*, spin-independent nonlinearities). However, the difference is not as pronounced as that observed in the experiment, leading us to conclude that while bosonic statistics may contribute as a secondary effect, it is not the primary mechanism driving spin-anisotropic interactions in double-layer samples. Therefore, a different mechanism is required to model the experimental results. Previous works have uncovered a dominant role of polarons[36,40] in causing the effective interactions between exciton-polaritons in transition metal dichalcogenide[41,42] and perovskite based samples[43,44]. While often considered in doped samples, this is not necessarily a requirement for polaron effects. The formation of such polarons can be described from the coupling of exciton-polaritons to phonons, which occurs via the Fröhlich interaction. For bare superlattice samples without the planar microcavity, the coupling strength between excitons and phonons is found to be similar in both ML and BL samples, as derived from the multimode Brownian oscillator (MBO) model (Section 6 of SM, Fig. S13 and Table 2). For two-dimensional polaritonic systems, the strength of this coupling $g^{2D}$ can be written as an overlap integral of two functions: the electrostatic potential created by optical phonons and the square of the envelope function of the electronic Bloch state[45]:

$$g^{LO}(q) = \int_{-\infty}^{+\infty} dz \chi^*(z) \phi_q(z) \chi(z)$$

(1)

Here $q$ is the in-plane wavevector, $\chi$ is the electronic envelope function and $\phi_q$ is the phonon induced electrostatic potential. The induced electrostatic potential can be estimated by considering Poisson's equation and solving it:

$$\phi_q(z) = C \int_{-\infty}^{+\infty} e^{ikz} f_q(k) \frac{q}{q^2+k^2} dk$$

(2)

where $f_q$ is an ansatz for the phonon induced polarization field and $k$ is the out-of-plane wavevector. Since the TMD layers are only a nanometer thick, compared to the scale of the microcavity, it is a good approximation to set $f_q(z) = \sum_{z_{TMD}} \delta(z - z_{TMD})$ where $z_{TMD}$ are the positions of the TMD layers. For a single layer, $\chi_{(z)}$ can be taken as a single Gaussian, while for a N2 it can be taken as two normalized Gaussians. The halfwidth of the Gaussians is chosen to be the thickness of the TMD layers. Calculating the Fröhlich interaction for various thickness of the $SiO_2$ buffer, we find that the exciton and phonon delocalization induces a decrease of the phonon-exciton coupling by up to a factor 5 for the bilayer with respect to the monolayer, as shown in Section 7 of SM, Fig. S14a.

When excitons are localized in single layers, the full breaking of translational symmetry in the growth direction allows them to couple to phonons with a wide range of wavevectors in the growth direction. However, in N2, as excitons have coupled to photons to form exciton-polaritons, they must have a symmetric phase distribution across the two layers. This assumption is also consistent with the calculation of the electric field profile of the ground state of the LPB, which shows that the electric field is indeed symmetric across the double-layer structure (Section 2 of SM). Only small differences (around 15% in intensity) are visible for extremely large spacer separations, as reported in Fig. S8. While momentum conservation is not fully enforced (as the excitons are still highly localized in the growth direction) they are no longer fully free to couple to phonons of any wavevector, favoring phonons that also have a symmetric phase distribution and a narrower range of wavevectors.

In order to account for this effect, we calculated the distribution of the phonon modes across the cavity layer using the transfer matrix method[46]. We consider the mirror/cavity/active layer design used in the experiment. Solving first for the eigenmodes and then for the eigenstates as explained in[46] we find that only a quarter of the phonon modes have the right symmetry to couple to the exciton-polariton in the case of the N2 while for the monolayer half do. An example of an asymmetric mode that does not couple in the case of a N2 is illustrated in Fig. 4c. See Methods for additional information. Therefore, one can write the ratio of the electron-phonon coupling for monolayer and N2 as

$$R = C_{ph}\left(\frac{g_{1L}^2}{g_{2L}^2}\right) \qquad (3)$$

where $C_{ph}$ is the ratio of the coupling phonon modes that we found to be 2 when comparing monolayer and N2. Combining the contribution from the delocalization of the phonon (Section 7 of SM, Fig. S15) and exciton densities and the symmetry condition, we find that the phonon contribution is one order of magnitude weaker for an N2 structure than for a monolayer one. The ratio R could be even further enhanced due to the higher localization of the electronic field in the monolayer case. This is reminiscent of the use of plasmonic modes to confine TMD exciton-polaritons to enhance their coupling to phonons[47].

Consequently, if we recall that in single TMD microcavities[44], the effective interactions are expected to be dominated by polarons, the same doesn't necessarily happen in double (or multi-layer) TMD microcavities as the electron-phonon interaction has been reduced by an order of magnitude. Therefore, as observed in our experiment, the effective interaction does not behave the same for monolayers and multilayers. For multilayer samples, we then expect that a different mechanism takes over, such as Coulomb interaction or phase-space filling[26,35], both resulting in spin-dependent interactions.

We believe our results are caused by a different exciton-phonon interaction strength in monolayers and multi-layered samples in coupled systems, which switches between a polaron-mediated spin-independent nonlinearity and spin-dependent nonlinearity. While this concept has not appeared in the literature on exciton-polaritons or excitons in TMDs, to our knowledge, it is reminiscent of a theoretical reduction of electron-phonon scattering that was discussed for double heterostructures previously[48]. While the concept was not pursued further for electrons in quantum well heterostructures, to our knowledge, it is effectively enhanced in our system due to the unique physics of exciton-polaritons where the presence of the photon forces a symmetry of the wavefunctions that only some phonons can couple to. However, the aforementioned explanation would imply an additional feature. Since the phonon electrostatic potential is evanescently decaying in the spacer layer (as shown by our transfer matrix calculations), there should be some critical spacer distance above which the phonons in the two layers become independent. In this case, no specific phase relation would be required between the layers, and one should return to an effective monolayer scenario. The optimum exciton-phonon coupling in TMDs occurs when the product of the phonon wavevector and exciton Bohr radius is on the order of $2^{49}$ and our transfer matrix calculations show that the phonon layers become essentially independent for a thick spacer for such wavevectors, as shown in Fig. 14b. This is also consistent with the experimental data shown in Fig. 3. In addition to the aforementioned factors, the superradiance effect where the collective oscillation of the dipoles maintains the phase coherence, exists in the bilayer system without the cavity[50]. In the presence of the photon, an analogy to this effect may also affect the strength of exciton-phonon coupling in the double-layer system. This rich interplay of spin-dependent interactions and spacer layer thickness underscores the advantage of utilizing multiple $WS_2$ layers in polariton systems over monolayers at room temperature for spin dependent polaritonic, while also providing an additional tuning knob.

**Layer number dependence**

Furthermore, we investigate the light-matter interactions of polaritons within a structure with four separated TMD layers (N4). An angle-resolved reflectivity map is shown in Figure 5a, which demonstrates that excitons in the multi-layer structure effectively coupled with cavity modes, resulting in the formation of exciton polaritons. The schematic diagram of the multi-layer sample is presented in Figure 5b. Each layer is separated by approximately 6nm, using the same fabrication process as employed for the N2 samples. A coupled oscillator model is used to fit the dispersion and calculate the vacuum Rabi splitting, which is approximately 72 meV for the N4 superlattice. The Rabi splitting exhibits a square root dependence on the number of embedded layers, aligning well with previously reported data. The blueshift of the ground state as a function of polariton density is shown in Figure 5b, which reveals clear spin-dependent interactions, highlighting the intricate interplay between spin dependent interaction and structural configuration in these multi-layer samples.

**Conclusions**

In summary, we demonstrate spin-anisotropic polariton nonlinearities in microcavities with multiple $WS_2$ layers, which was not observed in similar monolayer samples at room temperature. We propose that if polarons dominate the effective nonlinear response, then one must consider the strength of exciton-phonon coupling and this can be drastically different in monolayer and multilayer samples. While polarons contribute a spin independent nonlinear interaction in monolayers, the suppression of exciton-phonon coupling in multilayer samples suppresses polaron formation allowing other mechanisms such as spin selective Coulomb interaction and phase space filling to take over. In principle, this explanation also implies that if it were possible to suspend a single monolayer in air, the spin-anisotropic interaction could also be recovered, as phonon scattering would be greatly reduced in such a true two-dimensional geometry. Indeed, this would be an interesting topic of future work. Our results open up new and easy avenues for the manipulation of layer-spin coherence in 2D materials within cavity-engineered photonic systems. This information is particularly important for the design and development of spintronic devices based on TMD polaritons at room temperature.

**Acknowledgments**

J.Z., A.F., K.D., and T.C.H.L. gratefully acknowledge support from the Singapore Ministry of Education (MOE) Academic Research Fund Tier 3 grant (MOE-MOET32023-0003) "Quantum Geometric Advantage". Q. S., J. Z., W. G., and T.C.H.L. gratefully acknowledge support from the National Research Foundation project N-GAP (NRF2023-ITC004-001). Q.X. gratefully acknowledges strong funding support from the National Key Research and Development Program of China (Grant No. 2022YFA1204700), National Natural Science Foundation of China (No. 122507101126 and 12020101003), and support from the State Key Laboratory of Low-Dimensional Quantum Physics of Tsinghua University and the Tsinghua University Initiative Scientific Research Program. D.S., A.F. and V.A. gratefully acknowledge "Quantum Optical Networks based on Exciton-polaritons" (Q-ONE, N. 101115575, HORIZON-EIC-2022-PATHFINDER CHALLENGES EU project), "National Quantum Science and Technology Institute" (NQSTI, N. PE0000023, PNRR MUR project), and "Integrated Infrastructure Initiative in Photonic and Quantum Sciences" (I-PHOQS, N. IR0000016, PNRR MUR project). J.Z. gratefully acknowledges the Presidential Postdoctoral Fellowship support from the Nanyang Technological University.

**Author contributions:**

J.Z., A.F and T.L. conceived the ideas and designed the experiments. J.Z. prepared multi layered microcavity samples. J.Z. and A.F. carried out the optical spectroscopy measurements and analyzed data. K.D., K.S., Y.Z, and T.L. performed theoretical calculations. J.Z., A.F., K.D., D.S., Q.X., and T.L. wrote the manuscript with contributions from all authors. All authors discussed the results and commented on the manuscript.

**Competing interests:**

The authors declare that they have no competing interests.

**Data and materials availability:**

All data needed to evaluate the conclusions in the paper are present in the paper and/or the Supplementary Materials. Additional data related to this paper may be requested from the authors.

**Methods**

**Multilayer TMD microcavity fabrication**

In constructing the double and multi-layer $WS_2$ microcavities, we employed a configuration consisting of an optical cavity with a resonance wavelength ($\lambda_{Exc}/2$), sandwiched between two distributed Bragg reflectors (DBRs). The bottom DBR, comprising seven alternating layers of silicon dioxide ($SiO_2$) and titanium dioxide ($TiO_2$), and was deposited onto a sapphire substrate by using electron beam evaporation techniques (Cello, Ohmiker-50B). Following this, the first $SiO_2$ layer with thickness around ($\lambda_{Exc}/4$), was deposited via the same techniques. The double and multilayer TMD structures were engineered by interspersing each layer of $WS_2$, each separated by $SiO_2$ layers as spacer. Each TMD layer was mechanically exfoliated from high-quality, commercially available bulk $WS_2$ crystals (HQ Graphene) and identified through reflection spectroscopy. The assembly process involved isolating the first monolayer, which was then transferred to the bottom DBR using a dry polymer transfer technique. This was followed by the deposition of a $SiO_2$ spacer layer through e-beam evaporation. Next, the second monolayer was prepared on a polydimethylsiloxane (PDMS) substrate (PF-X4, Gel-Pack) and carefully aligned over the first ML with the assistance of a home-built transfer stage. With the PDMS layer removed, the next $WS_2$ layer was positioned on the $SiO_2$ layer, thus ensuring the alignment and structural integrity of the structure. The iterative application of this process enables the formation of multi $WS_2$ structures. Finally, the polymethyl methacrylate (PMMA) layer was spin-coated, followed by the careful placement of a transferred top DBR, comprising 6.5 pairs, thereby completing the fabrication of the superlattice microcavity.

**Optical spectroscopy**

In investigating the optical properties of microcavities, the reflectance spectra and maps were obtained using illumination from a spectrally broad tungsten-halogen light source. This light was focused into a 2 μm spot through 100X microscope objective with a numerical aperture of 0.9. Angle-resolved measurements were captured by focusing the back-focal plane of the objective through the narrow slit of a spectrometer's (Horiba, iHR550), a configuration complemented by a 600 lines/mm grating and interfaced with a 2D charge-coupled device (CCD) array (Horiba, Symphony II). The analysis of the upper and lower branches of the polariton dispersion, was refined by adjusting parameters such as Rabi splitting and exciton-photon detuning, which provided a robust framework for fitting the observed dispersion.

Further explorations into nonlinear optical phenomena were undertaken through a home-built transmission spectroscopy setup (Section 1 of SM Fig. S3). Our experiment utilized the 800 nm pulsed laser output from a Ti: Sapphire regenerative amplifier (characterized by a 1 kHz repetition rate and pulse durations approximately 100 fs in width). The pulses from this source were subsequently modified using an optical parametric amplifier (Spectra-Physics TOPAS), allowing for the generation of tunable pulses to resonantly excite the lower polariton branch (LPB).

**Theoretical calculations**

In Section 7 of SM (Fig. S15) is shown the phonon electrostatic potentials calculated from Eq. 2 for the monolayer sample and different N2 samples considered in the experiment. The number of modes is consistent with the number of layers accounted for in the calculation. The exciton-phonon coupling strength is essentially given by the overlap of the phonon electrostatic potential with the electron wavefunction, according to Eq. 1. Approximating the electron wavefunction as a pair of Gaussian functions centered on each layer, we can immediately rule out significant exciton-phonon coupling with some of the calculated phonon modes based on their symmetry. Modes for which the electrostatic potential is antisymmetric with respect to the centers of each TMD layer will not give significant coupling. Note, however, that similar modes exist in the monolayer, so this consideration actually does not change the fraction of phonon modes that are coupling. What is different between monolayer and N2 samples is that modes for which the phonon electrostatic potential is antisymmetric with respect to the center of the two TMD layers neither couple to electrons and they have no analogue for monolayers. Consequently, we can expect the exciton-phonon coupling strength (which is a combination of electron-phonon and hole-phonon coupling strengths) to be weaker in N2 by at least a factor of two. The actual coupling strength in N2 is further weakened, because of the delocalization of the exciton and phonon densities. In total, these two contributions lead to a coupling strength reduced by an order of magnitude in bilayer with respect to monolayer.

Having identified some of the modes as symmetric and some as antisymmetric, Fig. S14b shows the energy splitting between such pairs of modes for increasing space separation. The result is dependent on the phonon wavevector, where optimum exciton-phonon coupling is expected for $q * a_0 = 2$, where $a_0$ is the exciton Bohr radius[49]. We find that the energy splitting, which is representative of the strength of phonon coupling between TMD layers, decreases from meV range for small separations to near zero for 40 nm separation. Given that optical phonons in $WS_2$ and $SiO_2$ have linewidths around 0.1 meV and 0.2-0.6 meV, respectively[51,52], we conclude that before 40 nm separation the phonons in the two layers effectively decouple. In this case we can expect the behaviour of the system to return to that of individual monolayers. Indeed, this is consistent with the experiment, which finds that the C/L ratio (Fig. 4a) returns to a similar value to a monolayer above a critical distance.

**Figures and Tables**

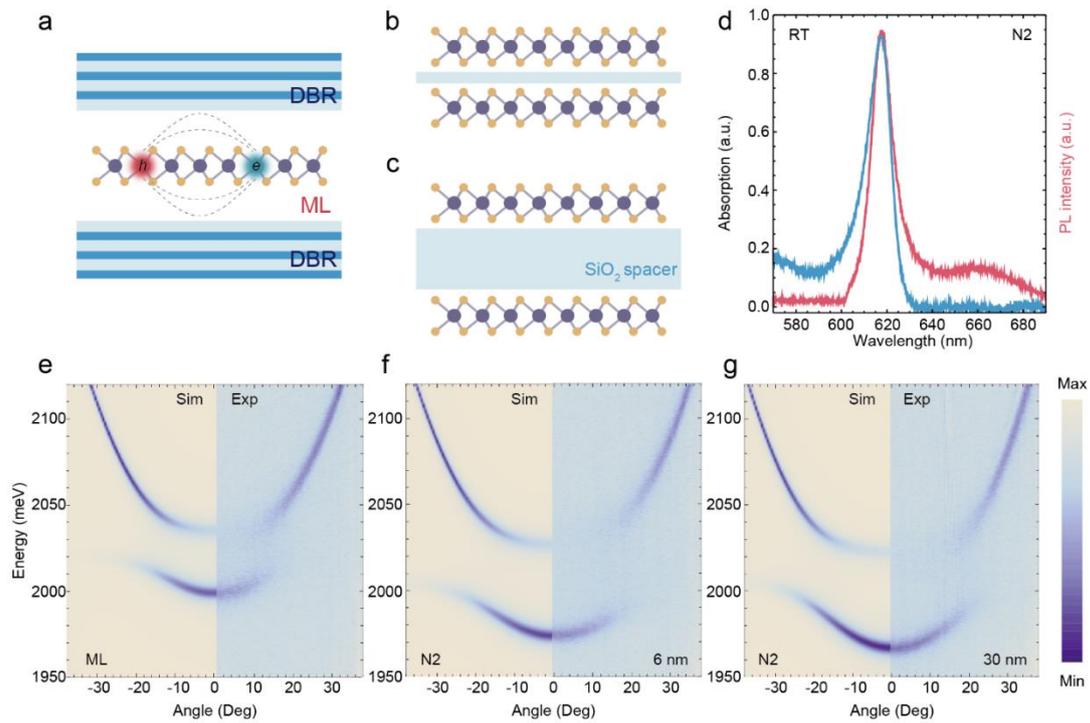

**Figure 1: Optical characterization of TMDs microcavities. a**, Schematic of the TMD based microcavities, consisting of monolayer WS$_2$ sandwiched between two distributed Bragg reflectors. **b, c,** Schematic of double layer WS$_2$ samples with different thickness of spacer layer and an arbitrarily twisted stack of monolayer WS$_2$, where SiO$_2$ spacers quench charge hopping. **d**, Absorption (ΔR/R, blue curve) and PL (red curve) spectra of double layer sample on the bottom DBR. **e, f, g,** Angle-resolved reflectivity map of monolayer (e), double layer sample with the spacer of 6 nm (f) and 30 nm (g), respectively. The left part is the Transfer Matrix Method calculation, and the right part is the experimental data, which are collected at RT by using a white light source.

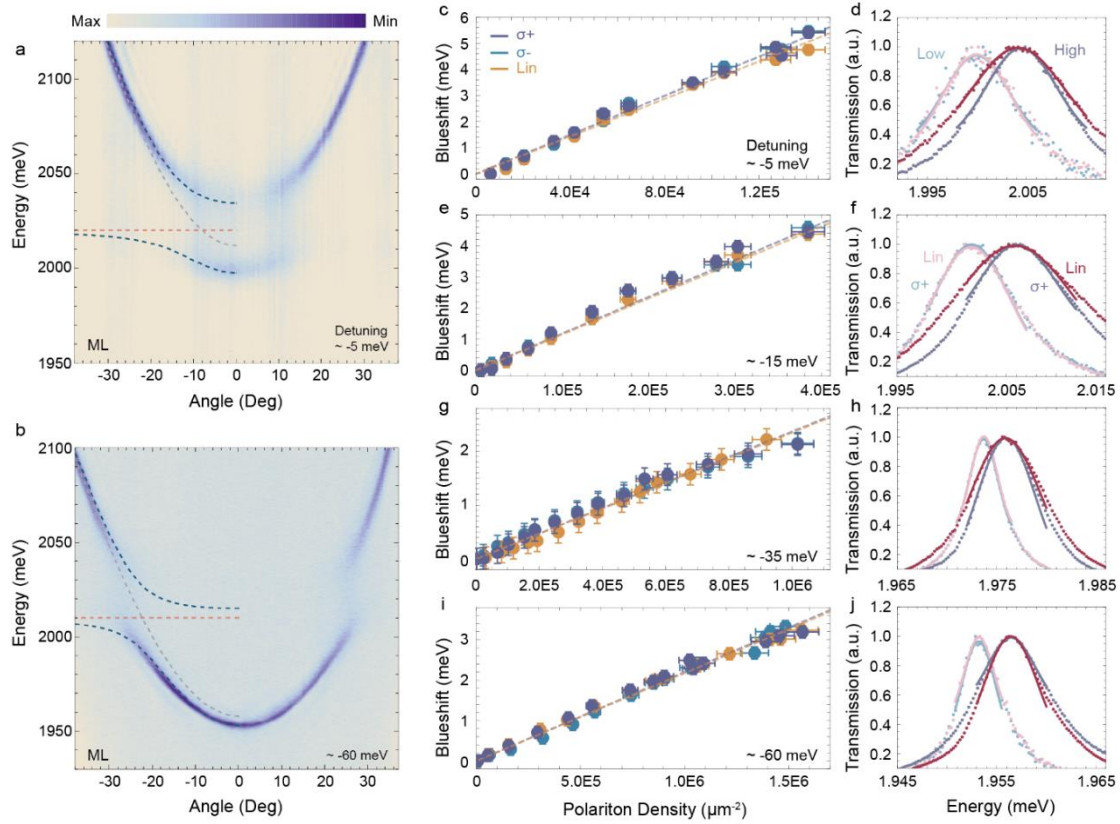

**Figure 2: Nonlinearities in monolayer TMD microcavity. a, b,** Angle-resolved reflectivity map of microcavities with different detuning values. Dashed lines are obtained from a coupled oscillator model and represent the cavity mode (gray dashed line), the excitonic resonance (red dashed line), the UPB and LPB (blue dashed line), respectively. **c, e, g** and **i**, The blueshift of the ground state of the LPB as a function of polariton density for linear (orange dots) and circular (dark purple for σ+ and dark blue dots for σ-) polarized excitations for the microcavities with different detuning values. The dashed orange and purple lines are linear fit of the experimental data. **d, f, h** and **j**, Normalized transmission spectra of the ground state of the LPB under resonant pumping at low (light color) and high density (dark color) for the linear (red) and circular (purple) polarized laser, with different detuning values. Experimental data points are represented by dots, while the solid lines indicate the Lorentzian fit function employed to extract the energy position.

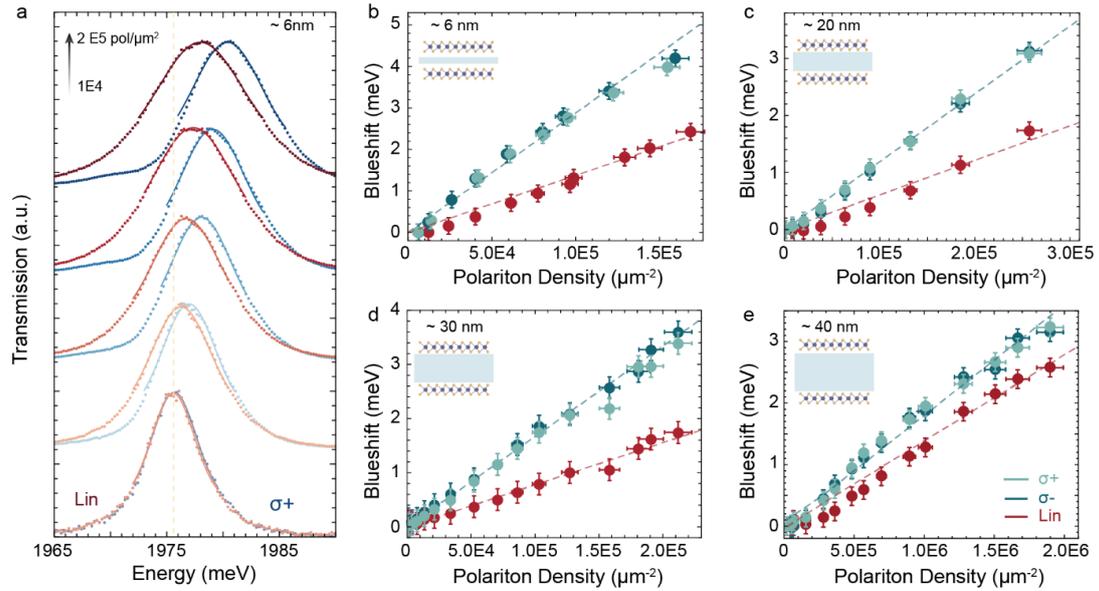

**Figure 3: The spin dependent nonlinearities in double layer WS$_2$ microcavities. a**, Normalized transmission spectra of the double layer microcavity (spacer of 6nm) with the circular (blue dots) and linear (red dots) polarized excitation of the LPB ground state. **b, c, d,** and **e,** Energy shift of the LPB ground state as a function of the polariton density for linear (red dots) and circular (light blue for σ+ and dark blue dots for σ-) polarized pumps and for different spacer thickness, 6 nm (**b**), 20 nm (**c**), 30 nm (**d**) and 40 nm (**e**), respectively, as indicated in the inset of each figure.

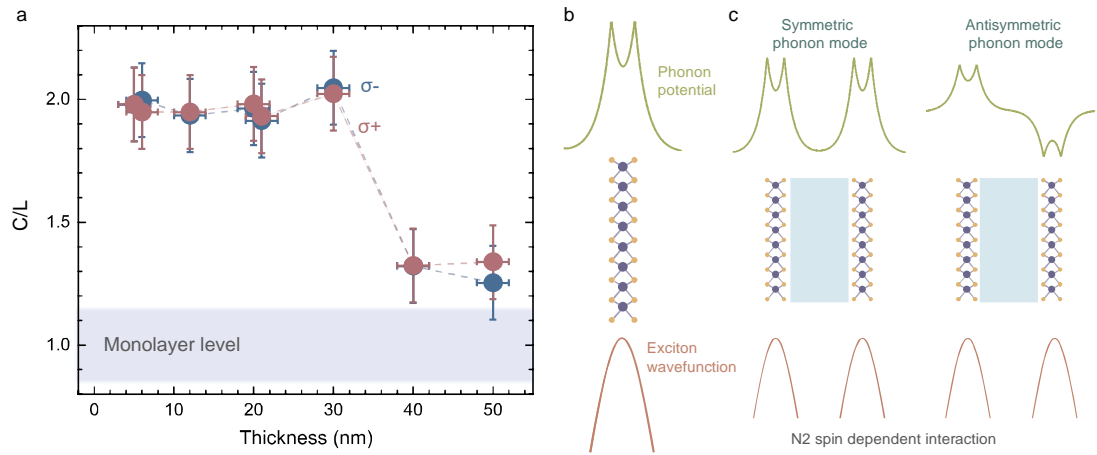

**Figure4: Circular/Linear slope ratio and theoretical model**. **a,** The experimentally extracted ratio of the blueshift slope, Circular/Linear polarized excitation, that is, a measure of spin-anisotropy of interactions, as a function of spacer layer thickness. We plot here the ratio obtained by using either a clockwise (blue points) or counterclockwise circular polarization (red points). The ratio is close to unity for a monolayer sample. Significant anisotropy is seen in the range of thicknesses 4-30nm, while above a critical separation between 30-40nm we return to a spin independent case. **b, c,** Optical phonons are not fully confined in the TMD layers and may penetrate into the surrounding $SiO_2$ material. The schematic forms of the phonon electrostatic potentials calculated with the transfer matrix technique are shown for monolayers (b) and a double layer (c) structure. As only symmetric modes in the double layer can couple to the symmetric exciton mode, the density of phonon states available for coupling is reduced. It is reduced further if one accounts for the delocalization of the exciton and phonon density effectively directly reducing the Frohlich coefficient.

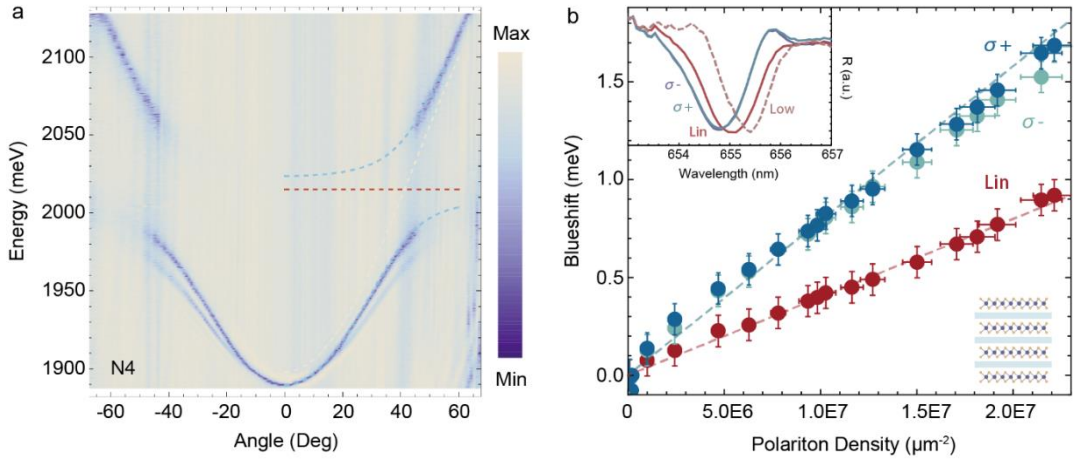

**Figure 5: The spin dependent nonlinearities in the multi-layer WS$_2$ sample. a,** Angle resolved reflectivity map obtained for the 4-layer (N4) sample. Dashed lines are obtained from a coupled oscillator model and represent the excitonic resonance (red dashed line), the UPB and LPB (blue dashed line), respectively. A Rabi splitting of around 72 meV is found. **b,** Energy shift of the LPB ground state as a function of the polariton density for linear (red dots) and circular (light blue for σ+ and dark blue dots for σ-) polarized pumps. The energy shift is extracted from fits of the reflection spectra. Error bars in the pump density represent uncertainty in the measured pump power and spot size. Error bars in blueshift represent uncertainty from the fitting procedure. The inset shows the reflectivity profiles at low power (red dashed line) and high power for linear (red solid line) and circular (light blue for σ+ and dark blue solid line for σ-) polarized pumps.